# ON THE ROLE OF THE BASIS OF MEASUREMENT IN QUANTUM GATE TELEPORTATION

F. V. Mendes, R. V. Ramos

fernandovm@deti.ufc.br    rubens@deti.ufc.br

*Lab. of Quantum Information Technology, Department of Teleinformatic Engineering – Federal University of Ceara - DETI/UFC, C.P. 6007 – Campus do Pici - 60455-970 Fortaleza-Ce, Brazil.*

Quantum teleportation is a powerful protocol with applications in several schemes of quantum communication, quantum cryptography and quantum computing. The present work shows the required conditions for a two-qubit quantum gate to be deterministically and probabilistically teleported by a quantum gate teleportation scheme using different basis of measurement. Additionally, we present examples of teleportation of two-qubit gates that do not belong to Clifford group as well the limitations of the quantum gate teleportation scheme employing a four-qubit state with genuine four-way entanglement.

## 1. Introduction

Since its proposal [1] quantum teleportation has played an increasing role in schemes of quantum communication, quantum cryptography and quantum computing [2-4]. In what concerns the teleportation of quantum gates (teleportation of the action of the quantum gate), a well known result is that gates belonging to Clifford group are deterministically teleported by a quantum teleportation scheme employing a pair of Bell states and Bell basis in the measurement, what results in error correction based on Pauli matrices [4,5]. This happens because Clifford group preserves Pauli group under conjugation. Although some works have generalized the teleportation of quantum states [6], including the usage of generic bases in the measurement [7], from the best of our knowledge, this has not been done for quantum gate teleportation. In this direction, the present work discusses the role of the basis of measurement in a two-qubit quantum gate teleportation scheme, showing explicitly the conditions to be satisfied by the quantum gate and the basis of measurement in order to have a successful teleportation. Additionally, we also give examples of teleportation of quantum gate that do not belong to Clifford group. At last, we show the result of the teleportation of a two-qubit quantum gate when a four-qubit state with genuine four-way entanglement is used.

The present work is outlined as follows: In Section 2 the role of the measurement basis on the teleportation of quantum states is discussed; in Section 3 the teleportation of quantum gates is reviewed; in Section 4 we describe the sufficient conditions for a two-qubit quantum gate to be teleported by a quantum teleportation scheme using a particular basis of measurement; Section 5 brings some examples of quantum gate teleportation with different basis, including examples of teleportation of quantum gates that do not belong to Clifford group; Section 6 discusses the quantum gate teleportation by a scheme employing a genuine four-way entangled state; at last, conclusions are drawn in Section 7.

## 2. Teleportation of Quantum States

The general circuit for teleportation of a single-qubit is shown in Fig. 1.

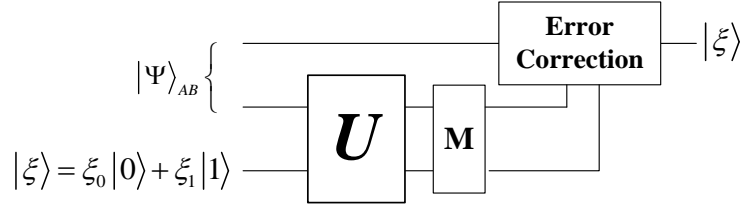

Fig. 1. General circuit for quantum teleportation of a single-qubit.

Basically, one qubit, let us say the second one, from the two-qubit state interacts with the single-qubit that will be teleported. After, measurements are realized and, according to the results obtained, a single-qubit error-correction is applied on the first qubit of the two-qubit state. If the teleportation is succeeded, the last qubit will be exactly in the same quantum state that the input qubit was. The quantum state evolution in Fig 1 is as follows:

$$(I \otimes U)|\Psi\rangle_{AB}|\xi\rangle_C = \sum_{i=1}^{4} \sqrt{p_i} V_i |\xi\rangle_A |\beta_i\rangle_{BC}. \tag{1}$$

In order to have a successful teleportation the condition $\langle \beta_i|\beta_j\rangle=\delta_{ij}$ must be obeyed. Let $|\phi\rangle$ be a general single-qubit state. Now, taking the inner product of $\langle\phi|\langle\beta_i|$ with (1) one gets

$$\langle\phi|\langle\beta_j|(I \otimes U_2)|\psi\rangle|\xi\rangle = \sum_{i=1}^{4} \sqrt{p_i} \langle\phi|V_i|\xi\rangle\langle\beta_j|\beta_i\rangle \tag{2}$$

$$\langle\phi|\langle\beta_j|(I \otimes U_2)|\psi\rangle|\xi\rangle = \sqrt{p_j} \langle\phi|V_j|\xi\rangle. \tag{3}$$

Using $|\psi\rangle=\psi_{00}|00\rangle+\psi_{01}|01\rangle+\psi_{10}|10\rangle+\psi_{11}|11\rangle$ in (3) it results in

$$\langle\phi|\langle\beta_j|(I \otimes U)(\psi_{00}|00\rangle|\xi\rangle+\psi_{01}|01\rangle|\xi\rangle+\psi_{10}|10\rangle|\xi\rangle+\psi_{11}|11\rangle|\xi\rangle)$$
$$= \sqrt{p_j}\langle\phi|V_j|\xi\rangle \tag{4}$$

$$\psi_{00}\langle\phi|0\rangle\langle\beta_j|U|0\xi\rangle+\psi_{01}\langle\phi|0\rangle\langle\beta_j|U|1\xi\rangle$$
$$+\psi_{10}\langle\phi|1\rangle\langle\beta_j|U|0\xi\rangle+\psi_{11}\langle\phi|1\rangle\langle\beta_j|U|1\xi\rangle = \sqrt{p_j}\langle\phi|V_j|\xi\rangle \tag{5}$$

$$\langle\phi|\begin{bmatrix}(\psi_{00}\langle\beta_j|U|0\xi\rangle+\psi_{01}\langle\beta_j|U|1\xi\rangle)|0\rangle \\ +(\psi_{10}\langle\beta_j|U|0\xi\rangle+\psi_{11}\langle\beta_j|U|1\xi\rangle)|1\rangle\end{bmatrix} = \sqrt{p_j}\langle\phi|V_j|\xi\rangle. \tag{6}$$

Substituting

$$|\xi\rangle = \xi_0|0\rangle + \xi_1|1\rangle \tag{7}$$

$$U|0\rangle|\xi\rangle = \xi_0 U|0\rangle|0\rangle + \xi_1 U|0\rangle|1\rangle \tag{8}$$

$$U|1\rangle|\xi\rangle = \xi_0 U|1\rangle|0\rangle + \xi_1 U|1\rangle|1\rangle \tag{9}$$

$$V_j = \begin{bmatrix} v_{11}^j & v_{12}^j \\ v_{21}^j & v_{22}^j \end{bmatrix} \tag{10}$$

in (6) one obtains

$$\langle\phi|\left\{\begin{bmatrix}(\psi_{00}\langle\beta_j|U|00\rangle+\psi_{01}\langle\beta_j|U|10\rangle)\xi_0+\\(\psi_{00}\langle\beta_j|U|01\rangle+\psi_{01}\langle\beta_j|U|11\rangle)\xi_1\end{bmatrix}|0\rangle+\begin{bmatrix}(\psi_{10}\langle\beta_j|U|00\rangle+\psi_{11}\langle\beta_j|U|10\rangle)\xi_0+\\(\psi_{10}\langle\beta_j|U|01\rangle+\psi_{11}\langle\beta_j|U|11\rangle)\xi_1\end{bmatrix}|1\rangle\right\}=$$
$$\langle\phi|\left[\sqrt{p_j}\left(v^j_{11}\xi_0+v^j_{12}\xi_1\right)|0\rangle+\sqrt{p_j}\left(v^j_{21}\xi_0+v^j_{22}\xi_1\right)|1\rangle\right]. \qquad(11)$$

Thus the $V_j$ matrix can be directly obtained from (11):

$$V_j=\frac{1}{\sqrt{p_j}}\begin{bmatrix}\psi_{00}\langle\beta_j|U_{00}\rangle+\psi_{01}\langle\beta_j|U_{10}\rangle & \psi_{00}\langle\beta_j|U_{01}\rangle+\psi_{01}\langle\beta_j|U_{11}\rangle\\ \psi_{10}\langle\beta_j|U_{00}\rangle+\psi_{11}\langle\beta_j|U_{10}\rangle & \psi_{10}\langle\beta_j|U_{01}\rangle+\psi_{11}\langle\beta_j|U_{11}\rangle\end{bmatrix}, \qquad(12)$$

where $U_{xy}=U|xy\rangle$ $x,y\in\{0,1\}$. Eq. (12) shows the relation between the two-qubit state used, the errors corrections required and the measurement basis used by Bob. It can be rewritten as

$$V_j=\frac{1}{\sqrt{p_j}}\psi\beta_U=\frac{1}{\sqrt{p_j}}\begin{bmatrix}\psi_{00} & \psi_{01}\\ \psi_{10} & \psi_{11}\end{bmatrix}\begin{bmatrix}\langle\beta_j|U_{00}\rangle & \langle\beta_j|U_{01}\rangle\\ \langle\beta_j|U_{10}\rangle & \langle\beta_j|U_{11}\rangle\end{bmatrix}. \qquad(13)$$

It is well known that teleportation is possible only if there is entanglement. This can be directly seen from (13). Since $V_j$ must belong to $U(2)$, one must have $|\det(V_j)|=1$. Hence $\det(\psi)=\psi_{00}\psi_{11}-\psi_{01}\psi_{10}\neq 0$. Note that $|\det(\psi)|^2$ is an entanglement measure (concurrence) for two-qubit pure states, hence the two-qubit state used cannot be disentangled. Now, let us consider the cases where the results obtained by Bob are equally probable, $p_j=1/4$, and the two-qubit state used is maximally entangled, $\psi_{00}=\psi_{11}=1/2^{1/2}$ and $\psi_{01}=\psi_{10}=0$. In this case Eq. (13) is simplified to

$$V_j=\sqrt{2}\begin{bmatrix}\langle\beta_j|U_{00}\rangle & \langle\beta_j|U_{01}\rangle\\ \langle\beta_j|U_{10}\rangle & \langle\beta_j|U_{11}\rangle\end{bmatrix}. \qquad(14)$$

It can be easily checked that, for a given $U$, when the measurement basis is $\{(|U_{00}\rangle\pm|U_{11}\rangle)/2^{1/2}, (|U_{01}\rangle\pm|U_{10}\rangle)/2^{1/2}\}$ the $V_j$ are the Pauli matrices. Note that a basis $\{|\beta_1\rangle,|\beta_2\rangle,|\beta_3\rangle,|\beta_4\rangle\}$ is useful for teleportation only if the matrices $V_1,..,V_4$ produced are unitary.

Now, let us consider $U=(H\otimes I)C_{\pi/8}$ (controlled-$\pi/8$) whose action in the canonical basis is given by

$$U_{x0}=U|x0\rangle=\left[|00\rangle+(-1)^x|10\rangle\right]\big/\sqrt{2} \qquad(15)$$

$$U_{x1}=U|x1\rangle=\left[\frac{(1+i)}{\sqrt{2}}\right]^x\left[|01\rangle+(-1)^x|11\rangle\right]\big/\sqrt{2}. \qquad(16)$$

Choosing $\{(|U_{00}\rangle\pm e^{i\pi/4}|U_{11}\rangle)/2^{1/2},(|U_{01}\rangle\pm e^{i\pi/4}|U_{10}\rangle)/2^{1/2}\}$ as measurement basis, the error correction is realized by the single-

qubit gates provided by Eq. (14) are $V_1^\dagger = [\pi/8]Z$, $V_2^\dagger = [\pi/8]$, $V_3^\dagger = XSZ$, $V_4^\dagger = XS$, where

$$\pi/8 = \begin{bmatrix} 1 & 0 \\ 0 & e^{i\pi/4} \end{bmatrix},\ S = \begin{bmatrix} 1 & 0 \\ 0 & i \end{bmatrix},\ Z = \begin{bmatrix} 1 & 0 \\ 0 & -1 \end{bmatrix},\ X = \begin{bmatrix} 0 & 1 \\ 1 & 0 \end{bmatrix}. \tag{17}$$

## 3. Teleportation of quantum gates

The general scheme for teleportation of two-qubit quantum gate is shown in Fig. 2.

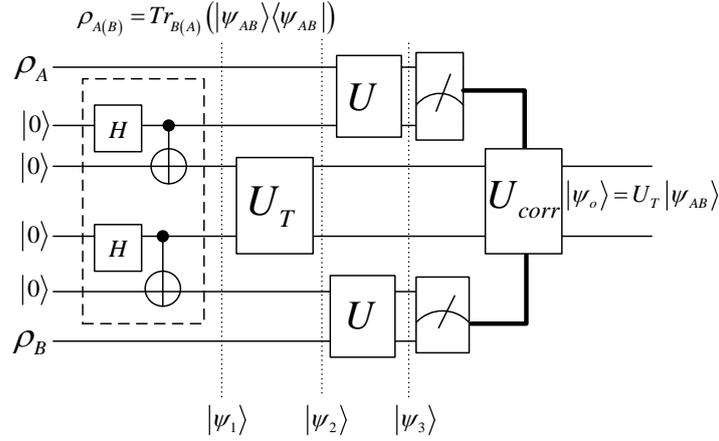

Fig. 2. General circuit for teleportation of the two-qubit quantum gate $U_T$.

Analyzing the quantum circuit in Fig. 2 for a successful teleportation, one has the following quantum state just before the measurements

$$|\psi_3\rangle = \frac{1}{2}\begin{bmatrix} \psi_{00}\begin{pmatrix} U|00\rangle U_T|00\rangle U|00\rangle + U|00\rangle U_T|01\rangle U|01\rangle \\ +U|01\rangle U_T|10\rangle U|00\rangle + U|01\rangle U_T|11\rangle U|01\rangle \end{pmatrix} \\ +\psi_{01}\begin{pmatrix} U|00\rangle U_T|00\rangle U|10\rangle + U|00\rangle U_T|01\rangle U|11\rangle \\ +U|01\rangle U_T|10\rangle U|10\rangle + U|01\rangle U_T|11\rangle U|11\rangle \end{pmatrix} \\ +\psi_{10}\begin{pmatrix} U|10\rangle U_T|00\rangle U|00\rangle + U|10\rangle U_T|01\rangle U|01\rangle \\ +U|11\rangle U_T|10\rangle U|00\rangle + U|11\rangle U_T|11\rangle U|01\rangle \end{pmatrix} \\ +\psi_{11}\begin{pmatrix} U|10\rangle U_T|00\rangle U|10\rangle + U|10\rangle U_T|01\rangle U|11\rangle \\ +U|11\rangle U_T|10\rangle U|10\rangle + U|11\rangle U_T|11\rangle U|11\rangle \end{pmatrix} \end{bmatrix}$$

$$= \frac{1}{4}\sum_{n,m=0}^{4} V_{nm} U_T \left[\psi_{00}|00\rangle + \psi_{01}|01\rangle + \psi_{10}|10\rangle + \psi_{11}|11\rangle\right]|\beta_n\rangle|\beta_m\rangle. \tag{18}$$

Now, taking the inner product of $\langle\beta_j|\langle\beta_k|$ with (18) one gets

$$\begin{bmatrix} \left( \langle\beta_j|U_{00}\rangle\langle\beta_k|U_{00}\rangle U_T\psi_{00}|00\rangle + \langle\beta_j|U_{00}\rangle\langle\beta_k|U_{01}\rangle U_T\psi_{00}|01\rangle \right) \\ + \left( \langle\beta_j|U_{01}\rangle\langle\beta_k|U_{00}\rangle U_T\psi_{00}|10\rangle + \langle\beta_j|U_{01}\rangle\langle\beta_k|U_{01}\rangle U_T\psi_{00}|11\rangle \right) \\ + \left( \langle\beta_j|U_{00}\rangle\langle\beta_k|U_{10}\rangle U_T\psi_{01}|00\rangle + \langle\beta_j|U_{00}\rangle\langle\beta_k|U_{11}\rangle U_T\psi_{01}|01\rangle \right) \\ + \left( \langle\beta_j|U_{01}\rangle\langle\beta_k|U_{10}\rangle U_T\psi_{01}|10\rangle + \langle\beta_j|U_{01}\rangle\langle\beta_k|U_{11}\rangle U_T\psi_{01}|11\rangle \right) \\ + \left( \langle\beta_j|U_{10}\rangle\langle\beta_k|U_{00}\rangle U_T\psi_{10}|00\rangle + \langle\beta_j|U_{10}\rangle\langle\beta_k|U_{01}\rangle U_T\psi_{10}|01\rangle \right) \\ + \left( \langle\beta_j|U_{11}\rangle\langle\beta_k|U_{00}\rangle U_T\psi_{10}|10\rangle + \langle\beta_j|U_{11}\rangle\langle\beta_k|U_{01}\rangle U_T\psi_{10}|11\rangle \right) \\ + \left( \langle\beta_j|U_{10}\rangle\langle\beta_k|U_{10}\rangle U_T\psi_{11}|00\rangle + \langle\beta_j|U_{10}\rangle\langle\beta_k|U_{11}\rangle U_T\psi_{11}|01\rangle \right) \\ + \left( \langle\beta_j|U_{11}\rangle\langle\beta_k|U_{10}\rangle U_T\psi_{11}|10\rangle + \langle\beta_j|U_{11}\rangle\langle\beta_k|U_{11}\rangle U_T\psi_{11}|11\rangle \right) \end{bmatrix} \cdot \frac{1}{2}$$

$$= \frac{1}{4} V_{jk} U_T \left[ \psi_{00}|00\rangle + \psi_{01}|01\rangle + \psi_{10}|10\rangle + \psi_{11}|11\rangle \right], \tag{19}$$

that can be rewritten as

$$U_T \beta_{jk} |\psi_{AB}\rangle = V_{jk} U_T |\psi_{AB}\rangle \tag{20}$$

$$\beta_{jk} = \sqrt{2} \begin{bmatrix} \langle\beta_j|U_{00}\rangle & \langle\beta_j|U_{10}\rangle \\ \langle\beta_j|U_{01}\rangle & \langle\beta_j|U_{11}\rangle \end{bmatrix} \otimes \sqrt{2} \begin{bmatrix} \langle\beta_k|U_{00}\rangle & \langle\beta_k|U_{10}\rangle \\ \langle\beta_k|U_{01}\rangle & \langle\beta_k|U_{11}\rangle \end{bmatrix} \tag{21}$$

$$|\psi_{AB}\rangle = \psi_{00}|00\rangle + \psi_{01}|01\rangle + \psi_{10}|10\rangle + \psi_{11}|11\rangle. \tag{22}$$

Equations (20)-(22) tell us that the quantum gate $U_T$ can be teleported only if the quantum gate $U$ and the measurement basis are chosen in such way that $U_T\beta_{jk}=V_{jk}U_T$ where, if the error correction are assumed to be local, $V_{jk}$ is decomposable in the tensor product of single-qubit gates. In other words, given $U$ and a measurement basis, there exist a specific set of quantum gates that can be teleported. For example, if $U=I$ and the measurement basis is the Bell basis, then we have the well known result that the error correction gates are Pauli matrices and the quantum gates belonging to Clifford group can be teleported.

Following the same procedure as before, for a two-qudit gate one has

$$U_T \beta |\psi_{AB}\rangle = V_{jk} U_T |\psi_{AB}\rangle \tag{23}$$

$$\beta = \sqrt{d} \begin{bmatrix} \langle\beta_j|U_{00}\rangle & \langle\beta_j|U_{10}\rangle & \cdots & \langle\beta_j|U_{(d-1)0}\rangle \\ \langle\beta_j|U_{01}\rangle & \langle\beta_j|U_{11}\rangle & \cdots & \langle\beta_j|U_{(d-1)1}\rangle \\ \vdots & \vdots & & \vdots \\ \langle\beta_j|U_{0(d-1)}\rangle & \langle\beta_j|U_{1(d-1)}\rangle & \cdots & \langle\beta_j|U_{(d-1)(d-1)}\rangle \end{bmatrix} \otimes \sqrt{d} \begin{bmatrix} \langle\beta_k|U_{00}\rangle & \langle\beta_k|U_{10}\rangle & \cdots & \langle\beta_k|U_{(d-1)0}\rangle \\ \langle\beta_k|U_{01}\rangle & \langle\beta_k|U_{11}\rangle & \cdots & \langle\beta_k|U_{(d-1)1}\rangle \\ \vdots & \vdots & & \vdots \\ \langle\beta_k|U_{0(d-1)}\rangle & \langle\beta_k|U_{1(d-1)}\rangle & \cdots & \langle\beta_k|U_{(d-1)(d-1)}\rangle \end{bmatrix} \tag{24}$$

$$|\psi_{AB}\rangle = \sum_{n,m=0}^{d-1} \psi_{nm}|nm\rangle, \tag{25}$$

while for $n$-qubit gate one obtains

$$U_T \beta |\psi_{12\ldots N}\rangle = V_{j_1\ldots j_N} U_T |\psi_{12\ldots N}\rangle \tag{26}$$

$$\beta = \prod_{j=1}^{N} \otimes \sqrt{2} \begin{bmatrix} \langle\beta_j|U_{00}\rangle & \langle\beta_j|U_{10}\rangle \\ \langle\beta_j|U_{01}\rangle & \langle\beta_j|U_{11}\rangle \end{bmatrix} \tag{27}$$

$$|\psi_{12\ldots N}\rangle = \psi_{00\ldots00}|00\ldots0\rangle + \psi_{00\ldots01}|00\ldots01\rangle + \psi_{00\ldots10}|00\ldots10\rangle + \psi_{11\ldots11}|11\ldots11\rangle. \tag{28}$$

## 4. Teleportation and separability of two-qubit gates

In this section we will discuss the conditions required for a basis to be useful for teleportation of a two-qubit gate or, alternatively, the conditions required for a gate to be teleported by a given basis. Firstly, let us assume the following notation: $\sigma_\mu = \mu$ and $\sigma_\upsilon = \upsilon$ are Pauli matrices, where $\mu,\upsilon \in \{I,X,Y,Z\}$. Furthermore, $\sigma_{\mu\upsilon} = \sigma_\mu \otimes \sigma_\upsilon$. Now, one has that

$$\left[\sigma_{\mu\mu},\sigma_{\upsilon\upsilon}\right]=\left[\sigma_{\mu\mu},\sigma_{I\mu}\right]=\left[\sigma_{\mu\mu},\sigma_{\mu I}\right]=\left[\sigma_{\mu I},\sigma_{I\upsilon}\right]=\left[\sigma_{I\mu},\sigma_{\upsilon I}\right]=0. \qquad (29)$$

As it is well know, a general single-qubit gate and a separable two-qubit gate can be represented, respectively, as

$$U_\Lambda = e^{-i\left(\frac{\lambda_1}{2}\sigma_Z\right)} e^{-i\left(\frac{\lambda_2}{2}\sigma_Y\right)} e^{-i\left(\frac{\lambda_3}{2}\sigma_Z\right)} \qquad (30)$$

$$U_{\Lambda\Omega} = \left[e^{-i\left(\frac{\lambda_1}{2}\sigma_Z\right)} \otimes e^{-i\left(\frac{\omega_1}{2}\sigma_Z\right)}\right]\left[e^{-i\left(\frac{\lambda_2}{2}\sigma_Y\right)} \otimes e^{-i\left(\frac{\omega_2}{2}\sigma_Y\right)}\right]\left[e^{-i\left(\frac{\lambda_3}{2}\sigma_Z\right)} \otimes e^{-i\left(\frac{\omega_3}{2}\sigma_Z\right)}\right]. \qquad (31)$$

Now, using

$$e^A \otimes e^B = e^{A \oplus B} \qquad (32)$$

$$A \oplus B = A \otimes I + I \otimes B \qquad (33)$$

in (31) one gets

$$U_{\Lambda\Omega} = e^{-i\left(\frac{\lambda_1}{2}\sigma_{ZI}+\frac{\omega_1}{2}\sigma_{IZ}\right)} e^{-i\left(\frac{\lambda_2}{2}\sigma_{YI}+\frac{\omega_2}{2}\sigma_{IY}\right)} e^{-i\left(\frac{\lambda_3}{2}\sigma_{ZI}+\frac{\omega_3}{2}\sigma_{IZ}\right)}. \qquad (34)$$

Now, according to KAK decomposition [8], a general two-qubit quantum gate is decomposed as

$$U = (U_A \otimes U_B) U_{NL} (U_C \otimes U_D) \qquad (35)$$

$$U_{NL} = e^{i(\theta_1 \sigma_{XX} + \theta_2 \sigma_{YY} + \theta_3 \sigma_{ZZ})} = e^{i(\theta_1 \sigma_{XX})} e^{i(\theta_2 \sigma_{YY})} e^{i(\theta_3 \sigma_{ZZ})}, \qquad (36)$$

where $U_A$, $U_B$, $U_C$ and $U_D$ are local single-qubit gates and $U_{NL}$ is the non-local part. Using (34) in (35)-(36), one obtains

$$U = e^{-i\left(\frac{\lambda_1}{2}\sigma_{ZI}+\frac{\omega_1}{2}\sigma_{IZ}\right)} e^{-i\left(\frac{\lambda_2}{2}\sigma_{YI}+\frac{\omega_2}{2}\sigma_{IY}\right)} e^{-i\left(\frac{\lambda_3}{2}\sigma_{ZI}+\frac{\omega_3}{2}\sigma_{IZ}\right)}$$
$$e^{i(\theta_1 \sigma_{XX})} e^{i(\theta_2 \sigma_{YY})} e^{i(\theta_3 \sigma_{ZZ})} \qquad (37)$$
$$e^{-i\left(\frac{\xi_1}{2}\sigma_{ZI}+\frac{\mu_1}{2}\sigma_{IZ}\right)} e^{-i\left(\frac{\xi_2}{2}\sigma_{YI}+\frac{\mu_2}{2}\sigma_{IY}\right)} e^{-i\left(\frac{\xi_3}{2}\sigma_{ZI}+\frac{\mu_3}{2}\sigma_{IZ}\right)}.$$

Considering again the two-qubit quantum gate teleportation, one has that teleportation with local error correction is possible if $U_T(\beta_j \otimes \beta_k)U_T^\dagger = V_j \otimes V_k$. Using the KAK decomposition of $U_T$, the left side can be rewritten as $[(U_A \otimes U_B)U_{NL}(U_C \otimes U_D)](\beta_j \otimes \beta_k)[(U_D^\dagger \otimes U_C^\dagger)U_{NL}^\dagger(U_B^\dagger \otimes U_A^\dagger)]$ hence, the teleportation is possible if $W = U_{NL}\beta_j' \otimes \beta_k' U_{NL}^\dagger$, where $\beta_j' \otimes \beta_k' = (U_C \otimes U_D)(\beta_j \otimes \beta_k)(U_D^\dagger \otimes U_C^\dagger)$, is separable. Using (34) and (36) $W$ can be written as

$$W = e^{i(\theta_1 \sigma_{XX})}e^{i(\theta_2 \sigma_{YY})}e^{i(\theta_3 \sigma_{ZZ})}e^{-i\left(\frac{\lambda_1}{2}\sigma_{ZI}+\frac{\omega_1}{2}\sigma_{IZ}\right)}e^{-i\left(\frac{\lambda_2}{2}\sigma_{YI}+\frac{\omega_2}{2}\sigma_{IY}\right)}e^{-i\left(\frac{\lambda_3}{2}\sigma_{ZI}+\frac{\omega_3}{2}\sigma_{IZ}\right)}e^{-i(\theta_3\sigma_{ZZ})}e^{-i(\theta_2\sigma_{YY})}e^{-i(\theta_1\sigma_{XX})}. \tag{38}$$

Using the commutation relations given in (29), one gets

$$\begin{aligned}W = &\; e^{i(\theta_1\sigma_{XX}+\theta_2\sigma_{YY})}e^{-i\left(\frac{\lambda_1}{2}\sigma_{ZI}+\frac{\omega_1}{2}\sigma_{IZ}\right)}e^{-i(\theta_1\sigma_{XX}+\theta_2\sigma_{YY})} \\ &\; e^{i(\theta_1\sigma_{XX}+\theta_3\sigma_{ZZ})}e^{-i\left(\frac{\lambda_2}{2}\sigma_{YI}+\frac{\omega_2}{2}\sigma_{IY}\right)}e^{-i(\theta_1\sigma_{XX}+\theta_3\sigma_{ZZ})} \\ &\; e^{i(\theta_1\sigma_{XX}+\theta_2\sigma_{YY})}e^{-i\left(\frac{\lambda_3}{2}\sigma_{ZI}+\frac{\omega_3}{2}\sigma_{IZ}\right)}e^{-i(\theta_1\sigma_{XX}+\theta_2\sigma_{YY})}.\end{aligned} \tag{39}$$

Therefore, $W$ is separable only if the unitary matrices

$$W_1 = e^{i(\theta\sigma_{\mu\mu})}e^{-i\left(\frac{\lambda}{2}\sigma_{ZI}\right)}e^{-i(\theta\sigma_{\mu\mu})} \tag{40}$$

$$W_2 = e^{i(\theta\sigma_{\mu\mu})}e^{-i\left(\frac{\lambda}{2}\sigma_{IZ}\right)}e^{-i(\theta\sigma_{\mu\mu})} \tag{41}$$

$$W_3 = e^{i(\theta\sigma_{\upsilon\upsilon})}e^{-i\left(\frac{\lambda}{2}\sigma_{YI}\right)}e^{-i(\theta\sigma_{\upsilon\upsilon})} \tag{42}$$

$$W_4 = e^{i(\theta\sigma_{\upsilon\upsilon})}e^{-i\left(\frac{\lambda}{2}\sigma_{IY}\right)}e^{-i(\theta\sigma_{\upsilon\upsilon})} \tag{43}$$

are also separable, where $\mu \in \{X,Y\}$ and $\upsilon \in \{X,Z\}$. One can find that the unitary matrices $W_1,\ldots,W_4$ are separable if the following condition is satisfied

$$\sin\left(\frac{\lambda}{2}\right)\sin(2\theta)\left[e^{i\frac{\lambda}{2}}\sin^2(\theta)+e^{-i\frac{\lambda}{2}}\cos^2(\theta)\right]=0. \tag{44}$$

The solutions of (44) are $(\theta,\lambda) \in \{(k\pi/2,x);(x,2k\pi);((2k+1)\pi/4,n\pi)\}$ where $x \in \Re$ and $k,n \in Z$. The first two solutions are not interesting because $\exp(i(k\pi/2)\sigma_{\mu\mu})$ is separable and $\exp(ik\pi\sigma_{\mu I}) = \exp(ik\pi\sigma_{I\mu})=I$. Therefore, the following theorem follows:

*Theorem* 1: Given a two-qubit quantum gate $U_T$ with KAK decomposition $U_T = (U_A \otimes U_B)U_{NL}(U_C \otimes U_D)$, and a two-qubit basis $\{|\beta_1\rangle, |\beta_2\rangle, |\beta_3\rangle, |\beta_4\rangle\}$, the quantum gate $U_T$ can be deterministically teleported if one of the following conditions is satisfied:

1)

$$U_{NL} = e^{i\left[\delta_x(2k_x+1)\frac{\pi}{4}\sigma_{XX} + \delta_y(2k_y+1)\frac{\pi}{4}\sigma_{YY} + \delta_z(2k_z+1)\frac{\pi}{4}\sigma_{ZZ}\right]}$$

$$(U_C \otimes U_D)(\beta_j \otimes \beta_k)(U_D^\dagger \otimes U_C^\dagger) = \begin{cases} e^{-i\left(\frac{n_1\pi}{2}\sigma_{ZI} + \frac{m_1\pi}{2}\sigma_{IZ}\right)} e^{-i\left(\frac{n_2\pi}{2}\sigma_{YI} + \frac{m_2\pi}{2}\sigma_{IY}\right)} e^{-i\left(\frac{n_3\pi}{2}\sigma_{ZI} + \frac{m_3\pi}{2}\sigma_{IZ}\right)} & \text{if } \delta_x = 1 \text{ or } \delta_x = 0, \delta_y = \delta_z = 1 \\ e^{-i\left(\frac{\lambda_1}{2}\sigma_{ZI} + \frac{\omega_1}{2}\sigma_{IZ}\right)} e^{-i\left(\frac{n_2\pi}{2}\sigma_{YI} + \frac{m_2\pi}{2}\sigma_{IY}\right)} e^{-i\left(\frac{\lambda_3}{2}\sigma_{ZI} + \frac{\omega_3}{2}\sigma_{IZ}\right)} & \text{if } \delta_x = \delta_y = 0, \delta_z = 1 \\ e^{-i\left(\frac{n_1\pi}{2}\sigma_{ZI} + \frac{m_1\pi}{2}\sigma_{IZ}\right)} e^{-i\left(\frac{\lambda_2}{2}\sigma_{YI} + \frac{\omega_2}{2}\sigma_{IY}\right)} e^{-i\left(\frac{n_3\pi}{2}\sigma_{ZI} + \frac{m_3\pi}{2}\sigma_{IZ}\right)} & \text{if } \delta_x = \delta_z = 0, \delta_y = 1 \end{cases}$$

2)

$$U_{NL} = e^{i\left[\frac{\pi}{4}\sigma_{XX} + \frac{\pi}{4}\sigma_{YY}\frac{\pi}{4}\sigma_{ZZ}\right]}$$

$$(U_C \otimes U_D)(\beta_j \otimes \beta_k)(U_D^\dagger \otimes U_C^\dagger) = e^{-i\left(\frac{\lambda_1}{2}\sigma_{ZI} + \frac{\omega_1}{2}\sigma_{IZ}\right)} e^{-i\left(\frac{\lambda_2}{2}\sigma_{YI} + \frac{\omega_2}{2}\sigma_{IY}\right)} e^{-i\left(\frac{\lambda_3}{2}\sigma_{ZI} + \frac{\omega_3}{2}\sigma_{IZ}\right)}$$

where

$$\beta_j \otimes \beta_k = \sqrt{2}\begin{bmatrix} \langle\beta_j|00\rangle & \langle\beta_j|10\rangle \\ \langle\beta_j|01\rangle & \langle\beta_j|11\rangle \end{bmatrix} \otimes \sqrt{2}\begin{bmatrix} \langle\beta_k|00\rangle & \langle\beta_k|10\rangle \\ \langle\beta_k|01\rangle & \langle\beta_k|11\rangle \end{bmatrix}$$

$$\begin{cases} \delta_{x,y,z} = 0, 1 \\ k_{x,y,z} = 0, \pm 1, \pm 2, ... \\ n_{1,2,3}, m_{1,2,3} = 0, \pm 1, \pm 2, ... \\ \lambda_{1,2,3}, \omega_{1,2,3} \in R \\ j, k = 1, 2, 3, 4 \end{cases}$$

Without loss of generality, we have considered $U = I$ in scheme in Fig. 2. The condition 1) comes from (44), while condition 2) comes from the action of the SWAP gate: $U_{swap}(U_x \otimes U_y)U_{swap}^\dagger = U_y \otimes U_x$ for any $U_x$ and $U_y$. Note that, if condition 1) is only partially satisfied (the tensor product is not satisfied for all values of $(j,k)$) then a probabilistic teleportation takes place. In this case the success probability is $N(j,k)/16$, where $N(j,k)$ is the number of pairs $(j,k)$ that satisfies the tensor product.

## 5. Examples of quantum gate teleportation with different bases

In this section we show some examples of teleportation with different bases. Firstly, let us consider the real basis $\beta_{ab}$ and its corresponding matrices $\beta_j$, $j=1,...,4$:

$$\beta_{ab} = \left\{ \begin{pmatrix} -a \\ b \\ b \\ a \end{pmatrix}, \begin{pmatrix} -b \\ a \\ -a \\ -b \end{pmatrix}, \begin{pmatrix} -a \\ -b \\ b \\ -a \end{pmatrix}, \begin{pmatrix} b \\ a \\ a \\ -b \end{pmatrix} \right\} \tag{45}$$

$$\beta_1 = \sqrt{2}\begin{bmatrix} a & -b \\ -b & -a \end{bmatrix}, \beta_2 = \sqrt{2}\begin{bmatrix} b & a \\ -a & b \end{bmatrix}, \beta_3 = \sqrt{2}\begin{bmatrix} a & -b \\ b & a \end{bmatrix}, \beta_4 = \sqrt{2}\begin{bmatrix} -b & -a \\ -a & b \end{bmatrix} \tag{46}$$

$$a^2 + b^2 = 1/2. \tag{47}$$

An example of such basis is

$$M_1 = \left\{ \frac{1}{2}\begin{pmatrix} -1 \\ 1 \\ 1 \\ 1 \end{pmatrix}, \frac{1}{2}\begin{pmatrix} -1 \\ 1 \\ -1 \\ -1 \end{pmatrix}, \frac{1}{2}\begin{pmatrix} -1 \\ -1 \\ 1 \\ -1 \end{pmatrix}, \frac{1}{2}\begin{pmatrix} 1 \\ 1 \\ 1 \\ -1 \end{pmatrix} \right\}. \tag{48}$$

The second basis to be considered is obtained taking the columns of the unitary matrix $U_{NL}$ given in (36):

$$\beta_{NL} = \left\{ \begin{pmatrix} \cos(\theta_1-\theta_2)e^{i\theta_3} \\ 0 \\ 0 \\ \sin(\theta_1-\theta_2)ie^{i\theta_3} \end{pmatrix}, \begin{pmatrix} 0 \\ \cos(\theta_1+\theta_2)e^{-i\theta_3} \\ \sin(\theta_1+\theta_2)ie^{-i\theta_3} \\ 0 \end{pmatrix}, \begin{pmatrix} 0 \\ \sin(\theta_1+\theta_2)ie^{-i\theta_3} \\ \cos(\theta_1+\theta_2)e^{-i\theta_3} \\ 0 \end{pmatrix}, \begin{pmatrix} \sin(\theta_1-\theta_2)ie^{i\theta_3} \\ 0 \\ 0 \\ \cos(\theta_1-\theta_2)e^{i\theta_3} \end{pmatrix} \right\} \tag{49}$$

$$(\theta_1,\theta_2) = \left\{ \begin{array}{l} \left(0,\pm\dfrac{\pi}{4}\right), \left(\pm\dfrac{\pi}{4},0\right), \left(0,\pm\dfrac{3\pi}{4}\right), \left(\pm\dfrac{3\pi}{4},0\right), \left(\pi,\dfrac{\pi}{4}\right), \left(\dfrac{\pi}{4},\pi\right) \\ \left(\pm\dfrac{\pi}{2},\pm\dfrac{\pi}{4}\right), \left(\pm\dfrac{\pi}{4},\pm\dfrac{\pi}{2}\right), \left(\pm\dfrac{\pi}{2},\pm\dfrac{3\pi}{4}\right), \left(\pm\dfrac{3\pi}{4},\pm\dfrac{\pi}{2}\right) \end{array} \right. \tag{50}$$

The angle $\theta_3$ can assume any real value. Note that taking values for $\theta_1$ and $\theta_2$ different from those values shown in (50) will result in a non valid basis since the matrices $\beta_j$ will not be unitary. Two examples of this class are the bases

$$M_{Bell} = \left\{ \frac{1}{\sqrt{2}}\begin{pmatrix} 1 \\ 0 \\ 0 \\ 1 \end{pmatrix}, \frac{1}{\sqrt{2}}\begin{pmatrix} 0 \\ 1 \\ 1 \\ 0 \end{pmatrix}, \frac{1}{\sqrt{2}}\begin{pmatrix} 1 \\ 0 \\ 0 \\ -1 \end{pmatrix}, \frac{1}{\sqrt{2}}\begin{pmatrix} 0 \\ 1 \\ -1 \\ 0 \end{pmatrix} \right\} \tag{51}$$

$$M_2 = \left\{ \frac{1}{\sqrt{2}}\begin{pmatrix} i \\ 0 \\ 0 \\ 1 \end{pmatrix}, \frac{1}{\sqrt{2}}\begin{pmatrix} 0 \\ -i \\ i \\ 0 \end{pmatrix}, \frac{1}{\sqrt{2}}\begin{pmatrix} 0 \\ 1 \\ 1 \\ 0 \end{pmatrix}, \frac{1}{\sqrt{2}}\begin{pmatrix} 1 \\ 0 \\ 0 \\ i \end{pmatrix} \right\}. \tag{52}$$

After some algebra, one can easily check that the CNOT can be deterministically teleported by the scheme in Fig. 2 if the basis $M_{Bell}$ is used, it can be probabilistically teleported if the basis $M_2$ is used and it cannot be teleported if the basis $M_1$ is used. Table 1 shows the probability of successful teleportation for a small set of gates when the basis $M_{Bell}$, $M_1$ and $M_2$ are considered.

| Gate | $M_{Bell}$ | $M_1$ | $M_2$ |
|---|---|---|---|
| CNOT | 1 | 0 | 0.5 |
| $C_{\pi/8}$ | 0.5 | 0 | 0.5 |
| CNOT$^{1/2}$ | 0.5 | 0 | 0.25 |
| SWAP$^{1/2}$ | 0.25 | 0.25 | 0.25 |
| $\exp(i\pi/4\sigma_{YY})$ | 1 | 1 | 0.25 |

Table 1. Probability of successful teleportation according to the basis used: $M_{Bell}$, $M_1$ and $M_2$.

Now, let us consider the quantum gate $T$ that is deterministically teleported by $M_{BELL}$ and $M_2$ but it is not teleported by $M_1$.

$$T = \begin{bmatrix} i & 0 & 0 & 0 \\ 0 & e^{i\phi} & 0 & 0 \\ 0 & 0 & e^{i\xi} & 0 \\ 0 & 0 & 0 & ie^{i(\phi+\xi)} \end{bmatrix}. \quad (53)$$

Considering the basis $M_2$ one has

$$\beta_1 = -iP = \begin{bmatrix} -i & 0 \\ 0 & 1 \end{bmatrix}, \beta_2 = \sigma_Y = \begin{bmatrix} 0 & -i \\ i & 0 \end{bmatrix}, \beta_3 = I = \begin{bmatrix} 1 & 0 \\ 0 & 1 \end{bmatrix}, \beta_4 = P\sigma_Z = \begin{bmatrix} 1 & 0 \\ 0 & -i \end{bmatrix} \quad (54)$$

and the results for $T\beta_j \otimes \beta_k T^\dagger$ shown in Table 2.

| $j$ | $k$ | $T\beta_j \otimes \beta_k T^\dagger$ | $j$ | $k$ | $T\beta_j \otimes \beta_k T^\dagger$ |
|---|---|---|---|---|---|
| 1 | 1 | $P \otimes -P$ | 3 | 1 | $V_\xi \otimes iP\sigma_Z$ |
| 1 | 2 | $P\sigma_Z \otimes -V_\phi \sigma_Z$ | 3 | 2 | $-V_\xi \sigma_Z \otimes iV_\phi$ |
| 1 | 3 | $P\sigma_Z \otimes iV_\phi$ | 3 | 3 | $V_\xi \sigma_Z \otimes -V_\phi \sigma_Z$ |
| 1 | 4 | $P \otimes P\sigma_Y \sigma_X$ | 3 | 4 | $-V_\xi \otimes P$ |
| 2 | 1 | $-V_\xi \sigma_Z \otimes P\sigma_Z$ | 4 | 1 | $P\sigma_Y \sigma_X \otimes P$ |
| 2 | 2 | $V_\xi \otimes V_\phi$ | 4 | 2 | $-V_\phi \sigma_Z \otimes iP$ |
| 2 | 3 | $V_\xi \otimes -V_\phi \sigma_Z$ | 4 | 3 | $P \otimes -V_\phi$ |
| 2 | 4 | $-V_\xi \sigma_Z \otimes iP$ | 4 | 4 | $P\sigma_Z \otimes P\sigma_Z$ |
| $V_{\phi,\xi} = \begin{bmatrix} 0 & -ie^{-i\phi,\xi} \\ ie^{i\phi,\xi} & 0 \end{bmatrix}$ | | | | | |

Table 2. Results of $T\beta_j \otimes \beta_k T^\dagger$ for teleportation of $T$ in (53) when basis $M_2$ is used.

Hence, the gate $T(\phi,\xi)$ is deterministically teleportable for any values for $\phi$ and $\xi$ when the basis $M_2$ is used (the same happens if $M_{Bell}$ is used). Two interesting cases are $T_1 = T(\phi = \pi/8, \xi = \pi/8)$ and $T_2 = T(\phi = \pi/7, \xi = \pi/13)$. They are deterministically teleportable but both of them do not belong to Clifford group. This is remarkable since up to now, from the best of our knowledge, none explicit example of a non-Clifford quantum gate teleportation was presented.

Another interesting situation is the two-qubit swap gate family: $(U_A \otimes U_B)\exp(i\pi/4(\sigma_{XX}+\sigma_{YY}+\sigma_{ZZ}))(U_C \otimes U_D)$. These gates are not separable but they do not change the entanglement of a two-qubit state. Some examples are:

$$SWAP = \begin{bmatrix} 1 & 0 & 0 & 0 \\ 0 & 0 & 1 & 0 \\ 0 & 1 & 0 & 0 \\ 0 & 0 & 0 & 1 \end{bmatrix}, Q = \begin{bmatrix} 0 & 0 & 0 & 1 \\ 0 & 1 & 0 & 0 \\ 0 & 0 & 1 & 0 \\ 1 & 0 & 0 & 0 \end{bmatrix}, R = \begin{bmatrix} 0 & 0 & 1 & 0 \\ 1 & 0 & 0 & 0 \\ 0 & 0 & 0 & 1 \\ 0 & 1 & 0 & 0 \end{bmatrix}. \tag{55}$$

According to condition 2 of Theorem 1, the swap family can always be deterministically teleported by any basis.

## 6. Quantum gate teleportation with a genuine four-way entangled state

In this section we consider the two-qubit quantum gate teleportation using a genuine four-way entangled state instead of a pair of Bell states. The four-qubit quantum state used is [9]

$$|\chi\rangle = (|0000\rangle - |0011\rangle - |0101\rangle + |0110\rangle + |1001\rangle + |1010\rangle + |1100\rangle + |1111\rangle)/8. \tag{56}$$

Following the steps described in (18)-(22), one gets the following quantum state before error correction

$$|\psi\rangle = \frac{\left(U_T U_1 \sigma_{XX}\beta_{jk}|\psi_{AB}\rangle + U_T U_1 \sigma_{ZZ}\beta_{jk}|\psi_{AB}\rangle\right)}{\sqrt{2}} \tag{57}$$

$$U_1 = \begin{bmatrix} 1 & 0 & 0 & 0 \\ 0 & 1 & 0 & 0 \\ 0 & 0 & -1 & 0 \\ 0 & 0 & 0 & 1 \end{bmatrix}. \tag{58}$$

Due to the superposition in (57), a successful teleportation of the two-qubit state $U_T|\psi_{AB}\rangle$ can never be achieved. However, using $U=U_T U_1$, (57) can be rewritten as

$$|\psi\rangle = \frac{\left(U\sigma_{XX}\beta_{jk}U^\dagger\right)U|\psi_{AB}\rangle + \left(U\sigma_{ZZ}\beta_{jk}U^\dagger\right)U|\psi_{AB}\rangle}{\sqrt{2}}, \tag{59}$$

what shows that a superposition of $U|\psi_{AB}\rangle$ with local errors is realized if $(U\sigma_{XX}\beta_{jk}U^\dagger)$ and $(U\sigma_{ZZ}\beta_{jk}U^\dagger)$ are separable. In particular, if Bell basis is used ($\beta_j$ is a Pauli matrix), $U$ belongs to Clifford group and $|\psi_{AB}\rangle = U^\dagger|00\rangle$, then one has

$$|\psi\rangle = \frac{\sigma_{nm}|00\rangle + \sigma_{rs}|00\rangle}{\sqrt{2}}. \tag{60}$$

In this case, the quantum gate teleportation schemes works as non–local quantum circuit whose input is $U^\dagger|00\rangle$ and the output is (probabilistically) one of the Bell states.

## 7. Conclusions

Summarizing, this work discussed: I) the sufficient conditions for a two-qubit quantum gate to be teleported by a quantum circuit having as resource two bell states and a given basis of measurement; II) the teleportation of gates out of Clifford group; III) the quantum gate teleportation using a four-qubit state with genuine four-way entanglement. From the work done, one can conclude that:

1. The Theorem 1 gives the sufficient conditions for a quantum gate $U_T$ to be teleported when a given basis is used.
2. From Theorem 1 one gets: 2.1) any Clifford gate has non-local part with angles $(2k+1)\pi/4$. 2.2) Any gate of the type $\sigma_{\mu\nu}\exp(i(2k_x+1)\pi/4\sigma_{XX}+ i(2k_y+1)\pi/4\sigma_{YY}+ i(2k_z+1)\pi/4\sigma_{ZZ}))\cdot\sigma_{\alpha\beta}$ belongs to Clifford group. 2.3) Any gate of the form $(U_A\otimes U_B)U_{NL}(U_R\otimes U_R)$ is teleported deterministically by the basis $|\beta_1\rangle,\ldots,|\beta_4\rangle$, where $|\beta_j\rangle = \sum_{x,y}\langle x|\beta_j|y\rangle/\sqrt{2}|xy\rangle$ and $\beta_j = U_R^\dagger \sigma_j U_R$. In this case, one can note that $\beta_j\beta_k = \beta_m$. Furthermore, it is possible to produce a continuum of gates out of Clifford group that are deterministically teleportable by the basis $|\beta_1\rangle,\ldots,|\beta_4\rangle$, and only probabilistically teleported by Bell basis, for example. In particular, if $U_1 = (U_R^\dagger \otimes U_R^\dagger)U_{NL}^1(U_R \otimes U_R)$ and $U_2 = (U_R^\dagger \otimes U_R^\dagger)U_{NL}^2(U_R \otimes U_R)$, then $U_1 \beta_{jk} U_1^\dagger = \beta_{nm}$ and $U_1 U_2 = (U_R^\dagger \otimes U_R^\dagger)(U_{NL}^1 U_{NL}^2)(U_R \otimes U_R)$. Hence, the set of matrices $(U_R^\dagger \otimes U_R^\dagger)U_{NL}^j(U_R \otimes U_R)$, with $j=1,2,3,\ldots$ labeling the different non-local parts, forms a group that preserves the group $\beta_j = U_R^\dagger \sigma_j U_R$ under conjugation.
3. The success probability of teleportation of a given gate depends on the basis used. In general one can say that a given basis has teleportation capability proportional to the amount of gates that can be teleported when it is used. In particular: 3.1) any basis that produces a unitary matrix $\beta_j=I$ (like the Bell basis), has teleportation capability larger than zero. In this case, the lowest success probability is 1/16; 3.2) any basis having at least one disentangled state has teleportation capability equal to zero: if the state $|\beta_j\rangle$ is disentangled, then $\det(\beta_j) = 0$ and $\beta_j$ is, obviously, not unitary.

## Acknowledgment

This work was supported by the Brazilian agency CNPq via Grant no. 303514/2008-6. Also, this work was performed as part of the Brazilian National Institute of Science and Technology for Quantum Information.

## References

1. C. H. Bennet, G. Brassard, C. Crépeau, R. Josza, A. peres, and W. K. Wooters (1993), *Teleporting an unknown quantum state via dual classical and Einstein-Podolsky-Rosen channels*, Phys. Rev. Lett., 70, 13, 1895.
2. Z. Zhan-Jun, L. Yu-Min, and M. Zhong-Xiao (2005), *Many-agent controlled teleportation of multi-qubit quantum information via quantum entanglement swapping*, Comm. in Theor. Phys., 44, 5, 847.
3. L. Chen, H. Lu, and W. Chen (2005), *Constructing a universal set of quantum gates via probabilistic teleportation*, Chin. Opt. Lett., 3, 4, 240.
4. D. Gottesman and I. L. Chuang (1999), *Quantum teleportation is a universal computational primitive*, Nature, 402, 390.
5. F. V. Mendes and R. V. Ramos (2011), *Schemes for teleportation of quantum gates*, Quant. Inf. Process., 10, 203.

6. S. Stenholm and P. J. Bardroff (1998), *Teleportation of N-dimensional states*, Phys. Rev. A, 58, 4373.
7. X.-Q. Xi, S.-R. Hao, B.-Y. Hou, and R.Hong Yue (2003), Quantum standard teleportation based on generic measurement bases, Comm. in Theor. Phys., 40, 415.
8. R. R. Tucci (2005), An introduction to Cartan's KAK decomposition for QC programmers, arXiv:quant-ph/0507171.
9. Y. Yeo, and W. K. Chua (2006), *Teleportation and dense coding with genuine multipartite entanglement*, Phys. Rev. Lett. 96, 060502/1-4.